\documentclass[conference]{IEEEtran}
\IEEEoverridecommandlockouts

\usepackage{cite}
\usepackage{amsmath,amssymb,amsfonts}

\usepackage{algorithmic}
\usepackage{graphicx}
\usepackage{textcomp}

\usepackage[ruled,vlined,linesnumbered]{algorithm2e}
\usepackage{subcaption}
\usepackage[hyphens]{url}
\usepackage[breaklinks=true]{hyperref}
\hypersetup{hidelinks=true, }
\usepackage[nameinlink]{cleveref}
\usepackage{fontawesome5}
\usepackage{stmaryrd}
\usepackage{wrapfig}
\usepackage{fontawesome5}

\usepackage{cancel}

\usepackage{multirow}
\usepackage[table,xcdraw]{xcolor}
\usepackage[normalem]{ulem}

\newboolean{showcomments}
\setboolean{showcomments}{true}
\ifthenelse{\boolean{showcomments}}
{ \newcommand{\mynote}[3]{
        \fbox{\bfseries\sffamily\scriptsize#1}
        {\small\textsf{\emph{\color{#3}{#2}}}}}}
{ \newcommand{\mynote}[3]{}}

\useunder{\uline}{\ul}{}
\def\BibTeX{{\rm B\kern-.05em{\sc i\kern-.025em b}\kern-.08em
    T\kern-.1667em\lower.7ex\hbox{E}\kern-.125emX}}
\begin{document}

\title{An Open-source Implementation and Security Analysis of Triad’s TEE Trusted Time Protocol
\thanks{This work was supported by a French government grant managed by the Agence Nationale de la Recherche under the France 2030 program, reference ``ANR-22-PEFT-0002'' as well as the ANR Labcom program, reference ``ANR-21-LCV1-0012''.
We thank the Complex Systems research group at the IIUN (University of Neuchâtel, Switzerland) for granting us access to their experimental platform.
}
}

\newboolean{anon}
\setboolean{anon}{false}
\newboolean{oneline}
\setboolean{oneline}{true}
\ifthenelse{\boolean{anon}}{
    \author{\IEEEauthorblockN{Anonymous Authors}
    Submission ID: XX
    }
}
{
    \ifthenelse{\boolean{oneline}}{
        \author{
            \IEEEauthorblockN{Matthieu Bettinger\text{*}\textsuperscript{§}, Sonia Ben Mokhtar\text{*}, Anthony Simonet-Boulogne\textsuperscript{\textdagger}}
            \IEEEauthorblockA{\text{*}INSA Lyon, CNRS, Universite Claude Bernard Lyon 1, LIRIS, UMR5205, 69621 Villeurbanne, France\\
            \{given-name\}.\{surname\}@liris.cnrs.fr \textsuperscript{§}Corresponding author}
            \IEEEauthorblockA{\textsuperscript{\textdagger}iExec Blockchain Tech, 69008 Lyon, France
            \{given-name\}.\{surname\}@iex.ec
            }
            Published in the \emph{2025 55th Annual IEEE/IFIP International Conference on} \\ \emph{Dependable Systems and Networks - Supplemental Volume (DSN-S)}
            \\ \url{https://doi.org/10.1109/DSN-S65789.2025.00053}
            \vspace{-0.25cm}
            }
    }
    {
        \author{\IEEEauthorblockN{Matthieu Bettinger}
        \IEEEauthorblockA{\textit{LIRIS-DRIM INSA Lyon} \\
        Lyon, France \\
        matthieu.bettinger@insa-lyon.fr}
        \and
        \IEEEauthorblockN{Sonia Ben Mokhtar}
        \IEEEauthorblockA{\textit{LIRIS-DRIM CNRS} \\
        Lyon, France \\
        sonia.ben-mokhtar@cnrs.fr}
        \and
        \IEEEauthorblockN{Anthony Simonet-Boulogne}
        \IEEEauthorblockA{\textit{iExec Blockchain Tech} \\
        Lyon, France \\
        anthony.simonet-boulogne@iex.ec}
        }
    }   
}

\newcommand{\sysname}{Triad\xspace}

\newcommand{\figpathtriadFmnoAEXaex}{imports/img/triad-F-12347noAEX-2025-03-24-21-41-47-aex.pdf}
\newcommand{\figpathtriadFmnoAEXdrift}{imports/img/triad-F-12347noAEX-2025-03-24-21-41-47-drift.pdf}
\newcommand{\figpathtriadFmnoAEXstatedurations}{imports/img/triad-F-12347noAEX-2025-03-24-21-41-47-state-durations.pdf}
\newcommand{\figpathtriadFmnoAEXstates}{imports/img/triad-F-12347noAEX-2025-03-24-21-41-47-states.pdf}
\newcommand{\figpathtriadFmnoAEXutnode}{imports/img/triad-F-12347noAEX-2025-03-24-21-41-47-ut-node.pdf}
\newcommand{\figpathtriadFmnoAEXutta}{imports/img/triad-F-12347noAEX-2025-03-24-21-41-47-ut-ta.pdf}
\newcommand{\figpathtriadFmallAEXaex}{imports/img/triad-F-allAEX-2025-03-24-21-49-23-aex.pdf}
\newcommand{\figpathtriadFmallAEXdrift}{imports/img/triad-F-allAEX-2025-03-24-21-49-23-drift.pdf}
\newcommand{\figpathtriadFmallAEXstatedurations}{imports/img/triad-F-allAEX-2025-03-24-21-49-23-state-durations.pdf}
\newcommand{\figpathtriadFmallAEXstates}{imports/img/triad-F-allAEX-2025-03-24-21-49-23-states.pdf}
\newcommand{\figpathtriadFmallAEXutnode}{imports/img/triad-F-allAEX-2025-03-24-21-49-23-ut-node.pdf}
\newcommand{\figpathtriadFmallAEXutta}{imports/img/triad-F-allAEX-2025-03-24-21-49-23-ut-ta.pdf}
\newcommand{\figpathtriadFpallAEXaex}{imports/img/triad-F+allAEX-2025-03-24-21-59-08-aex.pdf}
\newcommand{\figpathtriadFpallAEXdrift}{imports/img/triad-F+allAEX-2025-03-24-21-59-08-drift.pdf}
\newcommand{\figpathtriadFpallAEXstatedurations}{imports/img/triad-F+allAEX-2025-03-24-21-59-08-state-durations.pdf}
\newcommand{\figpathtriadFpallAEXstates}{imports/img/triad-F+allAEX-2025-03-24-21-59-08-states.pdf}
\newcommand{\figpathtriadFpallAEXutnode}{imports/img/triad-F+allAEX-2025-03-24-21-59-08-ut-node.pdf}
\newcommand{\figpathtriadFpallAEXutta}{imports/img/triad-F+allAEX-2025-03-24-21-59-08-ut-ta.pdf}
\newcommand{\figpathtriadffaexdelayshistogram}{imports/img/triad-ff-2025-03-24-22-18-44-aex-delays-histogram.pdf}
\newcommand{\figpathtriadffaex}{imports/img/triad-ff-2025-03-24-22-18-44-aex.pdf}
\newcommand{\figpathtriadffdrift}{imports/img/triad-ff-2025-03-24-22-18-44-drift.pdf}
\newcommand{\figpathtriadffstatedurations}{imports/img/triad-ff-2025-03-24-22-18-44-state-durations.pdf}
\newcommand{\figpathtriadffstates}{imports/img/triad-ff-2025-03-24-22-18-44-states.pdf}
\newcommand{\figpathtriadffutnode}{imports/img/triad-ff-2025-03-24-22-18-44-ut-node.pdf}
\newcommand{\figpathtriadffutta}{imports/img/triad-ff-2025-03-24-22-18-44-ut-ta.pdf}
\newcommand{\figpathtriadfflowinterraexdelayshistogram}{imports/img/triad-ff-low-interr-2025-03-24-22-53-49-aex-delays-histogram.pdf}
\newcommand{\figpathtriadfflowinterraex}{imports/img/triad-ff-low-interr-2025-03-24-22-53-49-aex.pdf}
\newcommand{\figpathtriadfflowinterrdrift}{imports/img/triad-ff-low-interr-2025-03-24-22-53-49-drift.pdf}
\newcommand{\figpathtriadfflowinterrstatedurations}{imports/img/triad-ff-low-interr-2025-03-24-22-53-49-state-durations.pdf}
\newcommand{\figpathtriadfflowinterrstates}{imports/img/triad-ff-low-interr-2025-03-24-22-53-49-states.pdf}
\newcommand{\figpathtriadfflowinterrutnode}{imports/img/triad-ff-low-interr-2025-03-24-22-53-49-ut-node.pdf}
\newcommand{\figpathtriadfflowinterrutta}{imports/img/triad-ff-low-interr-2025-03-24-22-53-49-ut-ta.pdf}

\newcommand{\figpathaddcount}{imports/img/count-2025-03-26-14-47-29-2-7-5-log.pdf}

\newcommand{\figpathdatamodel}{imports/img/TriHard - Data Model.pdf}
\newcommand{\figpathcommmodel}{imports/img/TriHard - Communication Model.pdf}
\newcommand{\figpathsysmodel}{imports/img/TriHard - System Model.pdf}
\newcommand{\figpaththrtsysmodel}{imports/img/TriHard - Threat-System Model.pdf}
\newcommand{\figpathbft}{imports/img/bftatt.png}
\newcommand{\figpatherror}{imports/img/clkerror.png}
\newcommand{\figpathfpattack}{imports/img/F+_attack.png}
\newcommand{\figpathfmattack}{imports/img/F-_attack.png}
\newcommand{\figpathfpmal}{imports/img/F+_1Mal.png}
\newcommand{\figpathfmmal}{imports/img/F-_1Mal.png}
\newcommand{\figpathfpmallow}{imports/img/F+_1Mal_low_AEX.png}
\newcommand{\figpathfattseqdiag}{imports/img/Fatt_seq_diag.png}

\maketitle

\begin{abstract}
    The logic of many protocols relies on time measurements. 
    However, in Trusted Execution Environments (TEEs) like Intel SGX, the time source is outside the Trusted Computing Base: a malicious system hosting the TEE can manipulate that TEE’s notion of time, e.g., jumping in time or affecting the perceived time speed. 
    Previous work like Triad propose protocols for TEEs to maintain a trustworthy time source. 
    However, in this paper, based on a public implementation of Triad that we contribute, we empirically showcase vulnerabilities to this protocol. 
    For example, an attacker controlling the operating system, and consequently the scheduling algorithm, may arbitrarily manipulate their local TEE’s clock speed. 
    What is worse, in case of faster malicious clock speeds, an attacker on a single compromised machine may propagate the attack to honest machines participating in Triad’s Trusted Time protocol, causing them to skip to timestamps arbitrarily far in the future.
    Then, infected honest machines propagate time-skips themselves to other honest machines interacting with them.
    We discuss protocol changes to Triad for higher resilience against such attacks.
\end{abstract}


\section{Introduction}

Provisioning trustworthy timestamps is critical for many applications, both for traditional and confidential computing (i.e., with integrity and confidentiality requirements fulfilled using Trusted Execution Environments, TEEs, like Intel SGX~\cite{Costan_Devadas_2016}).
Use-cases and impacts are far-reaching, ranging from TimeStamping Authorities~\cite{rfc2001TSA} and Proof-of-Elapsed-Time~\cite{Bowman_Das_Mandal_Montgomery_2021}; 
credential expiration and revocation~\cite{Alder_Scopelliti_VanBulck_Muhlberg_2023,Malhotra_Cohen_Brakke_Goldberg_2015};
latency-sensitive systems (e.g., trading~\cite{Addison_Andrews_Azad_Bardsley_Bauman_Diaz_Didik_Fazliddin_Gromoa_Krish_etal_2019}, real-time systems~\cite{Wang_Li_Li_Lu_Zhang_2022}, consistent and available databases~\cite{Corbett_Dean_Epstein_Fikes_Frost_Furman_Ghemawat_Gubarev_Heiser_Hochschild_etal_2013});
time-constrained resource allocation (e.g., resource leasing~\cite{trachTLeaseTrustedLease2021});
resilience to timeout manipulation (e.g., BFT leader changes, procrastinating BFT leaders~\cite{Aublin_Mokhtar_Quema_2013});
latency Quality-of-Service and resilience to malicious message delaying~\cite{bettinger2025cooltee};
to decorrelation between data and timestamps in time-series~\cite{nasrullahTrustedTimingServices2024}.

In the context of confidential computing, iExec~\cite{iexecwp} proposes a decentralized computing marketplace, where anyone can contribute datasets, applications, and TEE-enabled hardware.
These computing assets can then be matched by anyone to execute tasks, i.e., a given application's logic processing datasets while running on a server. 
TEEs on those servers are in place to guarantee execution integrity and confidentiality.
However, applications are user-defined and therefore arbitrary: they can rely on timestamps.
If those timestamps can be manipulated by a server's malicious owner, e.g., because they control the OS or hardware, then task results themselves can be manipulated, e.g., as part of use-cases presented above.

Indeed, with CPU-level TEEs, also called enclaves, like Intel SGX, the time source is outside the TCB: since 2020, the \texttt{sgx\textunderscore get\textunderscore trusted\textunderscore time} primitive~\cite{SGXgetTrustedTime} is deprecated.
Some previous trusted time mechanisms relied on this (e.g., TimeSeal~\cite{Anwar_Garcia_Han_Srivastava_2019}) or other deprecated primitives (e.g., on Intel TSX~\cite{Alder_Asokan_Kurnikov_Paverd_Steiner_2019,trachTLeaseTrustedLease2021}).
In recent years, new protocols have been proposed to address trusted time in Intel SGX, like T3E~\cite{hamidyT3EPracticalSolution2023a} and Triad~\cite{fernandezTriadTrustedTimestamps2023}.
While T3E is open-source~\cite{DistriNet/T3E_2024}, the original Triad protocol is closed-source (due to its implementation on top of the proprietary containerization solution Scone~\cite{Arnautov_2016}): 
we have been able to reach Triad's authors, but not to obtain implementation artifacts or supplementary specifications.
Therefore, in this paper, based on the specifications in the original Triad paper~\cite{fernandezTriadTrustedTimestamps2023}, we contribute a public implementation~\cite{TriHaRdcode} of Triad and reproduce results.
However, we also experimentally highlight attacks on the Triad protocol (the original paper does not include empirical evaluation of attack scenarios).
While Triad makes a cluster of TEEs collaborate to keep a shared notion of trusted time, assuming that all underlying OSs or hypervisors may be compromised,
we show that even a single compromised node can manipulate its own notion of time and cause honest nodes to skip to timestamps arbitrarily far in the future.
This effect then cascades to honest nodes contacting infected honest nodes, infecting them in turn.
Consequently, we also discuss protocol improvements towards a TEE trusted time protocol with higher resilience. 

This paper is structured as follows: 
first, \Cref{sec:related_work} presents related work on TEE trusted time mechanisms;
then, \Cref{sec:problem_statement} describes the Triad protocol and attacks upon it;
based on the contributed implementation of Triad, \Cref{sec:results} reproduces results in the original Triad paper and additionally illustrates the feasibility and impact of attacks on the Triad protocol;
finally, \Cref{sec:discussion} discusses protocol improvements to increase resilience to these attacks and \Cref{sec:conclusion} summarizes this work.
\section{Related work}\label{sec:related_work}

Time sources, like the TimeStamp Counter (TSC) on a CPU or a remote Time Authority (e.g., NTP time servers~\cite{NetworkTimeProtocol_1985,NTPsec}), are outside the TCB of TEEs.
Therefore, security mechanisms must be put in place to prevent timestamp tampering by an attacker, positioned at the hypervisor, OS, or on the network.
Hereafter, we describe recent solutions and their features.



\subsection{CPU-level TEE trusted time}

With Standard Intel SGX (SGX1), reads of the TSC with \texttt{rdtsc} are mediated by the OS, which can manipulate the value.
Meanwhile, with Scalable Intel SGX (SGX2), \texttt{rdtsc} instructions can be executed in-enclave, bypassing the OS (but remaining vulnerable to a malicious hypervisor).
Furthermore, a malicious OS controls the scheduling of enclaves, so enclaves may be interrupted arbitrarily.
Additions to Intel SGX like AEX-Notify~\cite{Constable_Bulck_Cheng_Xiao_Xing_Alexandrovich_Kim_Piessens_Vij_Silberstein_2023} enable the TEE to react to interruptions and handle them with arbitrary developer-defined code after the enclave resumes.
Triad~\cite{fernandezTriadTrustedTimestamps2023} makes a cluster of TEEs cooperate to keep a common and continuous notion of time.
Each TEE monitors its TSC and relies on AEX-Notify to detect when its notion of time continuity is severed: the TEE then either obtains a timestamp from a peer enclave in the cluster or, failing that, from a Time Authority (e.g., NTP time servers).
However, in this paper, we show vulnerabilities in Triad's calibration protocol and in communications between TEEs in the cluster, allowing time manipulations at single nodes and their propagation to others. 
T3E~\cite{hamidyT3EPracticalSolution2023a} uses a Trusted Platform Module~\cite{TPM20Library} as a time source, colocated with the TEE.
T3E hinders delaying messages coming from the TPM by limiting how many times the same timestamp can be used by the TEE and by stalling TEE execution if uses are depleted.
In turn, the underlying application will drop in throughput, which may be detected by that application's user.
However, quantifying appropriate numbers of uses, to neither block execution when there are no attacks nor give too much room for delay attacks, is complex, because effective ideal usage rates are code-, workload-, and hardware-dependent.
Further, if the application is non-interactive or, on the contrary, if there are many users consuming a remote TEE service, with some users who may be malicious, trustworthy monitoring of demanded and effective throughput is again difficult.
Additionally, the TPM can be configured by an attacker owning it (leading to up to a $\pm32.5\%$ drift-rate compared to real time~\cite{TPM20Library}) and may more generally be a target for attacks as a root of trust~\cite{Parno}.

\subsection{VM-level TEE trusted time}

VM-level TEEs like Intel TDX have started becoming available at Cloud Service Providers (e.g., Microsoft Azure~\cite{MicrosoftAzure2023}, Google Cloud~\cite{IntelGoogleCloud_2024}, IBM Cloud~\cite{RunIBMenclave_2025}).
With the TCB now comprising the operating system, attackers must devise new strategies to harm the system.
Notably, with respect to trusted time, TEEs like Intel TDX and AMD SEV-SNP have their time sources protected even against a malicious hypervisor, respectively with their virtualized TSC~\cite{tdxspecs2023} and SecureTSC~\cite{Neela} features.
With Intel TDX, writing in the TimeStamp Counter's registers is forbidden from inside the Trust Domain (TD), i.e., the guest VM. 
A hypervisor offsetting the TSC during a VM exit is similarly detected and results in an error upon VM entry~\cite{tdxspecs2023}.
Meanwhile, AMD's SecureTSC lets the hypervisor and VM guests modify the TSC without affecting other guests, whose TSC remains linearly increasing~\cite{Neela}.
However, VM-level TEEs' attack surface is still undergoing research~\cite{Neela,Gast_Weissteiner_Schröder_Gruss_2025,Wilke_Wichelmann_Rabich_Eisenbarth_2023,Mandal_Shukla_Mishra_Bhattacharya_Saxena_Mukhopadhyay_2025,Wilke_Sieck_Eisenbarth_2024} and a large TCB is more demanding to properly audit.
Our objective in this paper is therefore to get closer to the guarantees provided by VM-level trusted time mechanisms, but using CPU-level TEEs with a smaller TCB.
\section{Triad specifications \& attacks}\label{sec:problem_statement}

Now, we describe Triad~\cite{fernandezTriadTrustedTimestamps2023}, the attack vectors it originally aimed to mitigate, as well as our new attacks against Triad.

\subsection{Attacker model}

The attacker in Triad is assumed to control the operating system or hypervisor.
Notably, it can delay or drop any message between the TEE and other devices.
By controlling the OS, it may also arbitrarily cause interruptions.
Interestingly, while the original paper considers \emph{adding} interruptions, removing interruptions, e.g., by further isolating cores running Triad, is not mentioned.
We show in experiments how low interruption rates help strengthen some attacks.
Finally, regarding Triad's use of the TSC as a local time source, a hypervisor virtualizing the TSC may change its value's offset and scaling factor for the guest VM running a Triad node.

\subsection{Triad protocol and building blocks}

To protect against such an attacker, Triad uses the following building blocks.
First, to prevent arbitrary manipulations of the TSC, a thread in the TEE is dedicated to monitoring the TSC increment rate.
With SGX2, reading the TSC does not require exiting the enclave: as long execution remains in the enclave, the OS or hypervisor cannot manipulate the TSC read by the enclave.
Triad calibrates the monitoring thread by measuring increases in the TSC during uninterrupted executions of that thread. 
AEX-Notify~\cite{Constable_Bulck_Cheng_Xiao_Xing_Alexandrovich_Kim_Piessens_Vij_Silberstein_2023} enables the TEE to be aware of when such interruptions occur, called Asynchronous Enclave Exits (AEXs).
More precisely, arbitrary user logic can be triggered when a TEE thread resumes execution.

Once an AEX occurs, however, the timestamp is considered ``tainted'': an arbitrarily long time may pass before TEE execution resumes and the attacker may offset the TSC to make that duration seem shorter or even longer.
As a consequence, the TEE communicates with remote entities to refresh, to ``untaint'' its timestamp. 
The root of trust is a remote Time Authority (TA), e.g., an NTP server, which serves as the time reference.
Remote communications introduce network delays and the TEE is unavailable to client applications while its timestamp is tainted.
For shorter roundtrip delays and fewer requests to the TA, Triad nodes are organized in clusters of multiple TEEs.
After resuming from an interruption, a TEE first asks its peers in the cluster for a timestamp and only asks the TA upon failure to receive any responses from peers.

We now focus on two key protocol steps in Triad: TSC monitoring calibration and untainting using peer timestamps. 

\subsection{Attacking Triad's calibration protocol}

A critical aspect of Triad's calibration is to estimate the relationship between the passage of time with respect to the TA's reference clock and increments in the TSC.
To do so, Triad relies on roundtrip communications with the TA, bounded by the monitoring thread's continuous execution, i.e., between two AEXs. 
For example, the TEE may ask the TA to wait 1s before sending back the response.
Meanwhile, the monitoring thread checks its uninterrupted execution and reports the TSC increment once the TA's response arrives.

However, the TEE is not aware \emph{a priori} of how much reference time can pass between two AEXs and, as a consequence, cannot reliably bound the requested TA waittime by inter-AEX delays.
Even given such an estimate, e.g., using the TSC frequency measurement by the OS at boot-time, the effective network delay is also unknown by the TEE, giving an attacker a margin of freedom to delay messages.
The original paper's specifications do not fully define how calibration should be performed, besides repeated and independent short interactions with the TA which waits a requested amount of time $s$ before sending a response.
To account for the offset error introduced by network delays, we consider a linear regression over requested waittimes and measured TSC increments.
The slope is the TSC's increment rate with respect to the TA's reference time.
Without regression over multiple measurements, e.g., with only the mean obtained with long waittimes, the offset error would always overestimate the TSC's increment rate, i.e., slow the TEE's perceived clock speed.
In the original paper's experiments, some nodes do have positive drifts from the reference, i.e., their calibration does compensate the offset created by network delays. 
Note that Triad's measurements over short intervals can lead to poor precision in estimating the clock speed, even without attacks.
As a comparison, instead of measurements of around 1s, NTP uses long drift measurement timeframes, between 16s and 36h~\cite{NetworkTimeProtocol_1985}.

Based on the above considerations, we design the following ``F+'' and ``F--'' attacks that respectively increase and decrease a node $i$'s perceived TSC increment rate $F_{i}^{\text{calib}}$ compared to its real rate $F_{i}^{\text{TSC}}$, i.e., slow down or quicken the TEE's perceived passage of time.
A TEE, as part of its TSC speed calibration protocol, sends messages to the TA, which waits a requested duration $s$ included in the message.
Communications are authenticated and encrypted, so the attacker does not have access to $s$.
However, the attacker is able to measure network delays between its machine and the TA, as well as roundtrip times part of Triad's calibration protocol, so the attacker can estimate $s$.
To slow down the TEE's clock, the attacker causes a steeper regression, i.e., $F_{i}^{\text{calib}}>F_{i}^{\text{TSC}}$, by adding delays to messages with high $s$, which we call an F+ attack.
Conversely, in an F-- attack, i.e., with $F_{i}^{\text{calib}}<F_{i}^{\text{TSC}}$ and leading to a faster TEE clock, the attacker adds delays to messages with low $s$.

\subsection{Propagating the attack to TEE peers}

When the TEE is not calibrating, it asks its peers for timestamps upon resuming after an AEX.
If any peers are not also ``tainted'', they send their current timestamp.
In the original Triad protocol~\cite{fernandezTriadTrustedTimestamps2023}, the policy to handle peer timestamps is as follows:
if a received timestamp is higher than the local one (the last one before the interrupt), then the incoming timestamp becomes the new reference; 
otherwise, the local timestamp is increased by the smallest possible increment to ensure monotonicity when serving client applications.
Such a policy ensures that TEEs cannot go back in time.
However, it also means that all TEEs will follow the fastest clock in the cluster.
Such drift errors can persist, because the TA is only contacted if all TEEs are ``tainted'' at the same time.
\section{Results}\label{sec:results}

Based on these specifications, we implement the protocol in C++ for Intel SGX TEEs, as well as F+ and F-- attacks.
The source code is available~\cite{TriHaRdcode}.
All protocol communications use UDP and are encrypted using AES-256-GCM~\cite{Li_2020}.
In the provided implementation, TSC rate estimation is performed through regression over roundtrips of messages with 0s-sleep (immediate responses) and 1s-sleep at the TA. 

Research questions (RQ) in this paper are along two axes: 
\begin{itemize}
    \item[\emph{A.}] Can the proposed public implementation reproduce results in Triad's original paper? 
    \item[\emph{B.}] Given this implementation based on the original specifications, how resilient is Triad to presented attacks?
\end{itemize}

More precisely, reproducibility is explored according to the following questions:
\begin{itemize}
    \item[\emph{A.1}] How accurate is TSC-monitoring using a counter in an enclave thread?
    \item[\emph{A.2}] Without attacks, how available are Triad nodes and how much do they drift compared to a reference time source? 
\end{itemize}

Note that we can only compare results without attacks, as the original paper does not provide evaluation under attacks.
Next, we attack the Triad protocol ourselves and we evaluate the drift of nodes participating in the protocol, by launching F+ and F-- attacks from a single compromised Triad node.

\begin{figure*}
    \centering
    \begin{subfigure}{0.49\textwidth}
        \centering
        \includegraphics[trim={0mm 0mm 0 0mm}, scale=0.78]{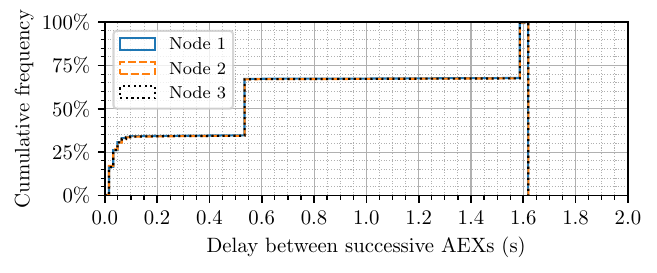}
        \subcaption{Triad-like~\cite{fernandezTriadTrustedTimestamps2023} simulated interruption distribution.}
        \label{fig:aexdelaytriad}
    \end{subfigure}
    \begin{subfigure}{0.49\textwidth}
        \centering
        \includegraphics[trim={0mm 0mm 0 0mm}, scale=0.78]{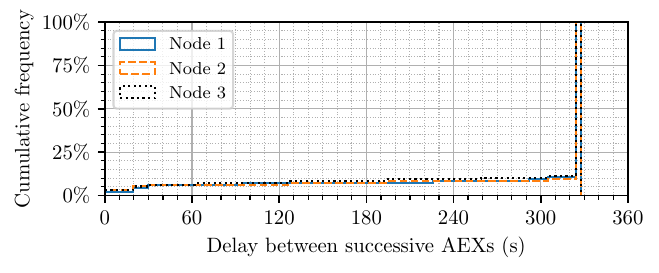}
        \subcaption{TSC monitoring core isolated from most OS interruptions.}
        \label{fig:aexdelaylowinterrupt}
    \end{subfigure}
    \caption{Cumulative distribution of delays between successive Asynchronous Enclave Exits (AEXs) on the TimeStamp Counter (TSC) monitoring enclave thread. 
    \Cref{fig:aexdelaytriad} is simulated on top of \Cref{fig:aexdelaylowinterrupt}'s system environment, by triggering AEXs at the TSC monitoring thread's core, using \texttt{rdmsr} instructions on that core's TSC MSR (Model Specific Register, \texttt{0x10} for TSC).}
    \label{fig:aexdelay}
\end{figure*}

For the following experiments, we run three Triad nodes and the TA on an Intel SGX2 (Scalable SGX) machine with 32 cores.
The TSC monitoring thread for each node is pinned to a core isolated from most OS interruptions.
As a result, those monitoring threads experience delays between AEXs as illustrated in \Cref{fig:aexdelaylowinterrupt}: most AEXs occur every 5.4 minutes.

To properly reproduce Triad's results, we simulate their distribution of inter-AEX delays (10ms, 532ms, and 1.59s, each with probability $1/3$), which we called ``Triad-like''.
The resulting distribution on our machine is shown in \Cref{fig:aexdelaytriad}, approximating Triad's original environment.
We do not have information on correlations that existed in their setup's successive delays between AEXs: we assume in this work that their successive delays were independent, i.e., $P(D_{i+1}=d)=P(D_{i+1}=d | D_{i})$, $\forall D_{i}, d$, with $D_{i}$ the duration between the $i^{\text{th}}$ and $(i+1)^{\text{th}}$ AEXs, and $d\in\{10,532,1590\}\text{ms}$.

Finally, to save space and avoid legends hiding data points, we do not repeat legends for every figure.
However, legends are consistent across figures, i.e., Nodes 1, 2, and 3 are always represented respectively in blue, orange, and black.
Note also that Nodes 1 and 2 are both honest in all experiments: Node 1's data points may overlap and hide Node 2's data points.

\subsection{Reproducing Triad~\cite{fernandezTriadTrustedTimestamps2023} results without attacks}

We now show our reproducibility results based on our public implementation~\cite{TriHaRdcode} of the Triad protocol.

\subsubsection{TSC monitoring with TEE enclave \texttt{INC}-counters}


With a fixed core-frequency, we run 10k measurements counting \texttt{INC} instructions until the TSC has incremented by 15E6, representing around 5ms in realtime (the TSC increments at $F^{\text{TSC}}=2899.999\text{MHz}$, as measured by the OS at boot-time).
Additionally, we have set the monitoring thread's core with the ``performance'' frequency-scaling governor (i.e., it runs at maximum frequency: 3500Mhz),
we obtain 632181\texttt{INC} as mean with 109.5\texttt{INC} of standard deviation.  
Removing two outliers (the experiment's first run with 621448\texttt{INC} and another with 630012\texttt{INC}), we obtain 632182\texttt{INC} as mean and go down to 2.9\texttt{INC} of standard deviation.
The range between measurement values without the two outliers is of 10\texttt{INC}: a monitoring thread running at a fixed frequency can therefore reliably detect TSC discrepancies, both in speed or time jumps (forward and back in time).
To answer \emph{RQA.1}, given that Intel CPUs allow only discrete pre-determined frequency settings~\cite{intelManual}, coupling this accurate but frequency-dependent monitoring with a less accurate but frequency-independent monitoring (e.g., memory-~\cite{fernandezTriadTrustedTimestamps2023} or randomness-~\cite{trachTLeaseTrustedLease2021} accesses) may lock an attacker from manipulating the TSC rate and offset.
However, we will show in later experiments that this mechanism is not sufficient to protect against an attacker manipulating the TEE's time perception: 
the attacker can still impact what duration of real elapsed time is equated to a number of TSC increments.

\subsubsection{Triad nodes availability and drift rates}

\begin{figure}
    \begin{subfigure}{\linewidth}
        \includegraphics[trim={-7mm 3mm 0 0mm}, scale=0.78]{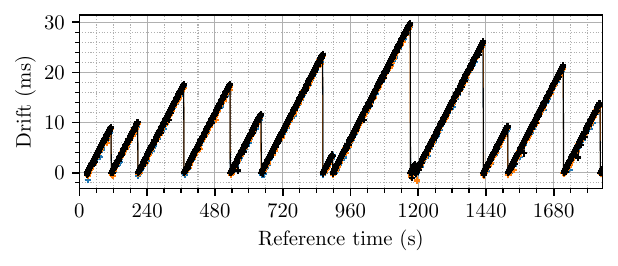}
        \subcaption{Clock drift per Triad node over time.}
        \label{fig:driftff}
    \end{subfigure}
    \begin{subfigure}{\linewidth}
        \includegraphics[trim={-7.5mm 3mm 0 0mm}, scale=0.78]{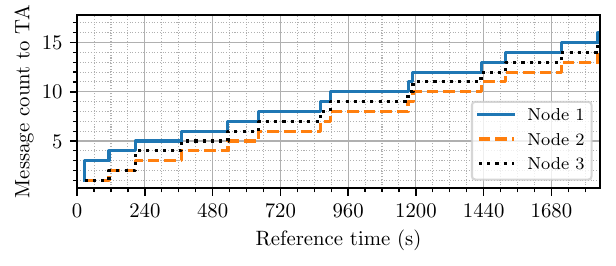}
        \subcaption{Number of received time references from the Time Authority.}
        \label{fig:taff}
    \end{subfigure}
    \caption{Long-term fault-free behavior of Triad nodes under \Cref{fig:aexdelaytriad}'s AEX delay distribution ($F_{1}^{\text{calib}}=2900.089\text{MHz}$; $F_{2}^{\text{calib}}=2900.113\text{MHz}$; $F_{3}^{\text{calib}}=2899.653\text{MHz}$).}
    \label{fig:ff}
\end{figure}

\begin{figure}
    \begin{subfigure}{\linewidth}
        \includegraphics[trim={-6.5mm 3mm 0 0mm}, scale=0.78]{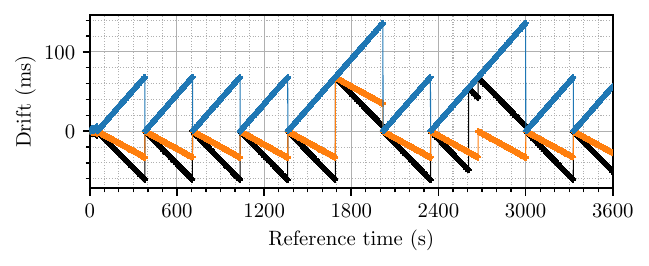}
        \subcaption{Clock drift per Triad node over time.}
        \label{fig:driftfflow}
    \end{subfigure}
    \begin{subfigure}{\linewidth}
        \includegraphics[trim={2mm 3mm 0 0mm}, scale=0.78]{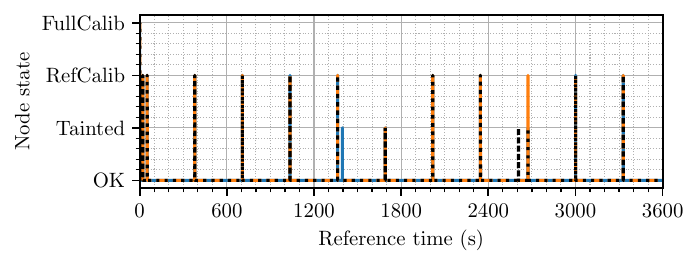}
        \subcaption{Timing diagram of Triad node states.}
        \label{fig:statesfflow}
    \end{subfigure}
    \caption{Long-term fault-free behavior of Triad nodes under \Cref{fig:aexdelaylowinterrupt}'s AEX delay distribution ($F_{1}^{\text{calib}}=2899.363\text{MHz}$; $F_{2}^{\text{calib}}=2900.260\text{MHz}$; $F_{3}^{\text{calib}}=2900.510\text{MHz}$).}
    \label{fig:fflow}
\end{figure}

Here, we assess Triad nodes' availability as well as drift rates they experience compared to the TA's reference time.
In all experiments, all nodes only required to perform full calibration, i.e., both clock time reference and speed, once with the TA.
\Cref{fig:statesfflow} illustrates this, showing the first hour of an experiment lasting 8h and a single stay in the ``FullCalib'' state at the start of the experiment.
The TSC was not manipulated during these experiments and monitoring cores ran at maximum frequency: 
no discrepancies in TSC update rates were detected by the TSC monitoring enclave thread's \texttt{INC}-instruction-counting.
Otherwise, additional full calibrations would have been triggered. 

Regarding drift, NTP's standard allowed clock drift-rate is 15ppm (parts-per-million, i.e., 15µs/s or 1.3s/day)~\cite{NetworkTimeProtocol_1985}.
In our scenarios without attacks, e.g., in \Cref{fig:driftff}, all nodes drift at around 110ppm (0.11ms/s), while Node 1 drifts at 210ppm in \Cref{fig:driftfflow}.
These \emph{effective} drift-rates are an order of magnitude higher than the standard \emph{upper bound} drift-rates.
This can be attributed to Triad's calibration protocol based on measurements over short durations, in the order of seconds or less, while standard clock synchronization like NTP monitor drift over long timeframes~\cite{NetworkTimeProtocol_1985}: $2^{\tau}\text{s}$, with $\tau\in\llbracket 4,17\rrbracket$, i.e., 16s to 36h.
Clock drifts reset to 0 in \Cref{fig:driftff} when a node's message count to the TA increments in \Cref{fig:taff}, i.e., when the node calibrates its time reference with the TA in absence of peer responses.
Figures in the original paper do not allow estimating their drift rates.

Nodes are unavailable to serve timestamps if they are tainted or calibrating: for \Cref{fig:ff}'s 30min experiment without attacks, each node's availability to serve timestamps exceeds 98\% including initial calibration.
\Cref{fig:fflow}'s experiment over 8 hours shows that availability can rise to 99.9\%. 

In a low-AEX setting, monitoring cores experience only a few AEXs with minutes of delay between them, as previously shown in \Cref{fig:aexdelaylowinterrupt}.
However, in our OS setup, these specific remaining interruptions by the OS do not target select individual cores but rather all cores.
This means that because all three nodes run on the same machine's cores, even under simulated AEXs, their TSC monitoring threads sometimes experience an AEX simultaneously (with higher probability than the original Triad experiment setup).
As a result, their timestamps will become tainted at the same time and they will not be able to fetch a fresh one from their peers: they must contact the TA, resetting their drifts.
This behavior also explains the sawtooth pattern of each node's drift time-series in \Cref{fig:driftff}.

Compared to \Cref{fig:ff}, without those correlated simultaneous AEXs, we can expect the node $i$ which underestimates the TSC frequency $F_{i}^{\text{calib}}$ the most to lead all other nodes to drift positively, i.e., to perceive a faster passage of time, even without attacks.
A low-AEX environment helps showcase this behavior, i.e., in \Cref{fig:fflow} at reference times $t=1705\text{s}$ for Node 2 and $t\in\{1705,2623,2688\}\text{s}$ for Node 3.
See how in \Cref{fig:statesfflow}, those nodes do not perform a time reference calibration with the TA (``RefCalib''), but were able to switch from a tainted to an OK state using peer untainting.
This results in time jumps for Nodes 2 and 3 to Node 1's increased clock time, i.e., in \Cref{fig:driftfflow}, by 50--70ms, followed by the nodes resuming timestamp updates at their own clock speeds.

To summarize and answer \emph{RQA.2}, Triad nodes exhibit high availability to serve timestamps.
However, even without attacks, a node underestimating $F^{\text{TSC}}$ is able to lead all other nodes to follow its drift, e.g., in \Cref{fig:driftff}, Node 3's drift (we have $F_{3}^{\text{calib}}<F^{\text{TSC}}<F_{1}^{\text{calib}}<F_{2}^{\text{calib}}$).
Furthermore, without simultaneous AEXs, this can happen arbitrarily long.

\subsection{Triad~\cite{fernandezTriadTrustedTimestamps2023} under attacks}

Hereafter, we launch F+ and F-- attacks on a single Triad node among the three and observe the system's behavior.

\subsubsection{Node 3 launching an F+ attack}

\begin{figure}
    \includegraphics[trim={2mm 3mm 0 0mm}, scale=0.78]{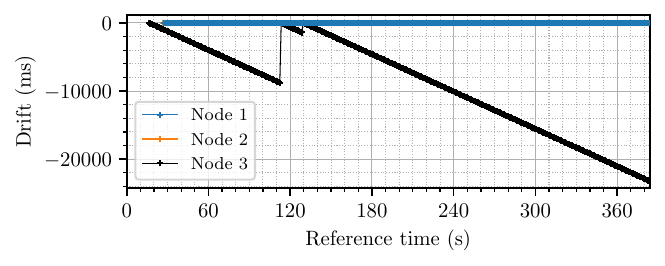}
    \caption{Clock drift of Triad nodes under an F+ attack on Node 3 ($F_{3}^{\text{calib}}=3191.224\text{MHz}$; $F_{1}^{\text{calib}}=2900.223\text{MHz}$; $F_{2}^{\text{calib}}=2900.595\text{MHz}$), which is in \Cref{fig:aexdelaylowinterrupt}'s low AEX environment, while Nodes 1 and 2 experience \Cref{fig:aexdelaytriad}'s Triad-like AEXs.}
    \label{fig:driftfpone}
    \vspace{-0.25cm}
\end{figure}

To start, we slow down the perceived time speed at Node 3 using an F+ attack, adding a 100ms delay to the TA's 1s-sleep messages.
In \Cref{fig:driftfpone}, Node 3 is additionally set in a low-AEX environment.
Besides two calibrations with the TA (due to correlated simultaneous AEXs on all cores), Node 3 drifts at -91ms/s (-9.1E4ppm).

\begin{figure}
    \includegraphics[trim={-2mm 3mm 0 0mm}, scale=0.78]{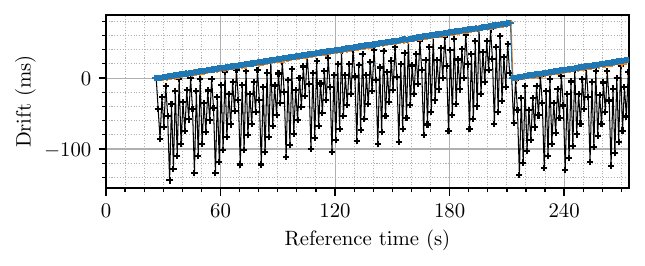}
    \caption{Clock drift of Triad nodes under an F+ attack on Node 3 ($F_{3}^{\text{calib}}=3191.210\text{MHz}$; $F_{1}^{\text{calib}}=2898.751\text{MHz}$; $F_{2}^{\text{calib}}=2900.836\text{MHz}$), with all nodes experiencing \Cref{fig:aexdelaytriad}'s Triad-like AEXs.}
    \label{fig:driftfpall}
\end{figure}

In \Cref{fig:driftfpall}, Node 3 experiences Triad-like AEXs.
The calibrated frequency is nearly the same as in the previous case (with a 4E-6 relative difference).
Node 3's drift now oscillates between two bounds: Nodes 1 and 2's drifts when it obtains peer timestamps after an AEX; and -150ms when it updates timestamps based on its own slow clock in-between AEXs.

\subsubsection{Node 3 launching an F-- attack}

\begin{figure}
    \begin{subfigure}{\linewidth}
        \includegraphics[trim={-1mm 3mm 0 0mm}, scale=0.78]{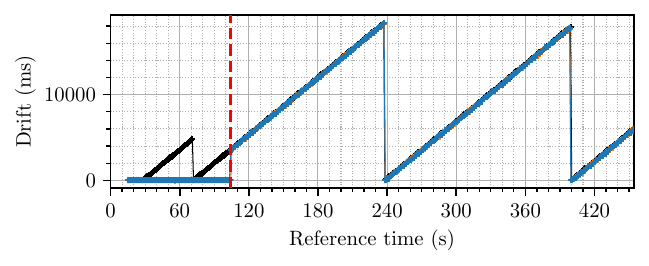}
        \subcaption{Clock drift per Triad node over time.}
        \label{fig:driftfmall}
    \end{subfigure}
    \begin{subfigure}{\linewidth}
        \includegraphics[trim={-5mm 3mm 0 0mm}, scale=0.78]{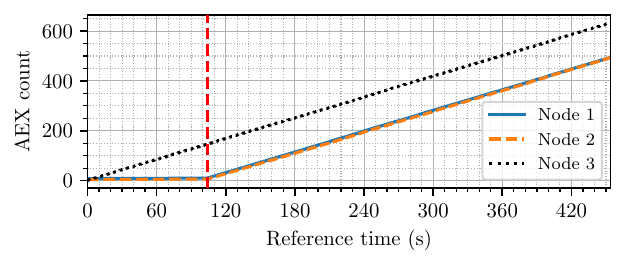}
        \subcaption{Number of AEX over time.}
        \label{fig:aexfmall}
    \end{subfigure}
    \caption{Triad nodes behavior under an F-- attack on Node 3 ($F_{3}^{\text{calib}}=2609.951\text{MHz}$; $F_{1}^{\text{calib}}=2899.347\text{MHz}$; $F_{2}^{\text{calib}}=2900.052\text{MHz}$), which experiences \Cref{fig:aexdelaytriad}'s Triad-like AEXs, while Nodes 1 and 2 experience \Cref{fig:aexdelaylowinterrupt}'s low AEXs for reference time $t<104\text{s}$, then Triad-like ones for $t\geq104\text{s}$, i.e., after the dashed red line.}
    \label{fig:fmall}
\end{figure}

Now, we quicken perceived time speed at Node 3 using an F-- attack, adding a 100ms delay to TA's immediate (0s-sleep) messages.
Node 3 drifts positively at 113ms/s.
To better highlight that attack's impact, Nodes 1 and 2 start the experiment with rare AEXs, then, after 104s (dashed red line in \Cref{fig:fmall}), they experience Triad-like AEXs.
\Cref{fig:aexfmall} shows the total number of AEXs at each node over time: while Node 3's AEXs linearly increase from the start, Nodes 1 and 2's AEXs stay around 0 for $t<104\text{s}$ then also linearly increase.
\Cref{fig:driftfmall} illustrates the nodes' drift before and after the change in AEX rates.
Without AEXs, both Nodes 1 and 2 experience relatively low drift-rates, similar to the case without attacks.
However, with higher AEX rates once $t>104\text{s}$, both nodes now communicate with the compromised Node 3 and use its timestamps, because they are farther in time than theirs.
As a result, both nodes jump forward in time, e.g., by around 35ms at $t=104\text{s}$, then alternate between their own clock's timestamps between AEXs and forward time jumps after AEXs.  

Finally, to answer \emph{RQB}, we observe that Triad is particularly sensitive to F-- attacks speeding up a single TEE's perceived passage of time, because that TEE propagates its positive drift to TEE peers.
With frequent AEXs, slowing down time with an F+ attack is less impactful, causing only the compromised node to oscillate between its peers' timestamps and its own slow clock measurements.
However, an attacker controlling the OS can prevent AEXs altogether, causing an arbitrary negative drift to that compromised node.
Additionally, these attacks on a node's calibration do not negatively affect availability: AEXs dictate how often the node communicates with peers and the TA.
In fact, as a consequence, a lower AEX rate, for example used above to strengthen an F+ attack, increases availability.

\section{Discussion}\label{sec:discussion}

In this section, we discuss our empirical results and propose protocol changes to address existing vulnerabilities.

First, in the original protocol, events to refresh a TEE's timestamp only come from outside the TCB, from attacker-controlled OS interruptions.
A compromised node may use its miscalibrated clock speed arbitrarily long.
To reduce the attack power, an in-TCB trigger can be added, e.g., using deadlines after given numbers of TSC increments.
As before, when a deadline is reached, the enclave will try to check its timestamp's quality or obtain a better one.

Moreover, the base protocol's synchronization precision and accuracy for uncompromised nodes should be close to that of a system without attacks.
With Triad~\cite{fernandezTriadTrustedTimestamps2023}, nodes with honest OSs instead use a calibration protocol designed to restrain an attacker's power, with a cost in synchronization quality.
Triad's calibration phases with short-duration measurements of clock speed and offset can be replaced by more mature synchronization protocols like NTPsec~\cite{NTPsec}. 
If honest, uncompromised nodes exist, they will be able to calibrate high-quality clocks over time.
Standard synchronization protocols use the notion of consistency in (sub-)sets of clocks in a system~\cite{Owicki}.
Given clocks with timestamps $t_{i}$ and $e_{i}$ an estimation on each clock's possible drift error, consistent subsets of clocks have all their intervals $t_{i} \pm e_{i}$ overlap with a non-empty intersection.
These clocks are usually called true-chimers.
Additionally, this method can be applied to clock time and speed.

With such a synchronization protocol, a node may now check if its clock is consistent with the TA.
If the node is compromised, messages may be delayed by the attacker: it may be consistent with a time reference offset towards the past.
However, honest nodes communicating with the compromised node will not consider it a true-chimer.
Nodes may publish, e.g., on a blockchain, or simply to other nodes, their list of true-chimers.
Nodes with the highest timestamp obtained from the TA have the most credibility to be honest.
Furthermore, many secure distributed protocols already rely on the assumption of an honest (super-)majority.
Under such an assumption, a majority clique of true-chimers may be used to maintain clock consistency and rely less often on the TA.

Ongoing work is dedicated to implementing and evaluating a protocol that builds upon these specifications.
\section{Conclusion}\label{sec:conclusion}

Many traditional and confidential computing protocols rely on trustworthy provisioning of timestamps.
However, current TEEs like Intel SGX lack a built-in trusted time mechanism.
We contribute a public implementation of the recent but closed-source Triad protocol, which aims to provide trusted time to a cluster of Intel SGX TEEs.
We show vulnerabilities in the protocol, e.g., that enable a single compromised node to propagate faster passage of time to honest nodes.
Finally, we discuss the found vulnerabilities' sources and propose changes for higher resilience against such attacks.
Future work is dedicated to a protocol that implements those changes.
%
%
%
\bibliographystyle{IEEEtran}
\bibliography{imports/bibliography}
%
\end{document}